# Reconstructing Implicit Scientific Knowledge: Evaluating LLM Agents through End-to-End Reproduction of Astronomy Studies

*[Yuehui Wang][1]*, [Xinyu Qi][1], [Guirong Xue][1] , [Cheng Wang][1], [Yangbin Xie][1] ,[Xiaoyu Tang][1] ,[Cong Sun][1] ,*

1 [Research Center for Scientific Data Hub, Zhejiang Lab, Hangzhou, China]

*Correspondence: [wangyuehui@zhejianglab.org]

## Abstract

The integration of large language models (LLMs) into scientific workflows is accelerating, yet their ability to reconstruct the reasoning underlying published research remains unexplored. Papers specify explicit procedures while leaving many methodological dependencies—data selection, calibration corrections, priors, and domain assumptions—implicit. This ambiguity complicates the evaluation of LLM-based agents, since a failure to reproduce a result may reflect either limitations of the agent or underspecification in the source. We present a framework that evaluates agents through end-to-end reproduction, separating execution from verification and computational failure from methodological ambiguity. We apply it to fourteen astronomy studies: a case study from The Astrophysical Journal and thirteen papers published in Nature. Eleven of the thirteen contained an ambiguity preventing a uniquely specified reproduction path. In a controlled case study, twelve predefined paths — a 3x2x2 sensitivity analysis over sample definition, sky masking, and parallax zero-point treatment—gave estimates from 2.16 to 3.53 kpc for the same quantity, with only one recovering the published value (about 2.70 kpc). The published value was never used as an optimization target, selection criterion, or stopping condition; the matching path was found only after all twelve had run. Crucially, the decisive information (a +0.02 mas parallax zero-point correction) was already in the paper, but the agents did not recognize its causal relevance until the analysis made the effect visible. Matching a published outcome therefore does not validate reconstruction of the underlying reasoning, and the bottleneck is as often a failure to connect relevant information as to retrieve it. End-to-end reproduction thus serves both as a test of reproducibility and as a framework for evaluating implicit scientific knowledge in AI systems.

# 1 Introduction

Large language models (LLMs) are increasingly capable of performing individual components of scientific work, including literature retrieval, code generation, data analysis, mathematical reasoning, and the interpretation of computational results (Vaswani et al., 2017; Brown et al., 2020; OpenAI, 2023). Recent agentic systems extend these capabilities by allowing models to invoke external tools, execute programs, retrieve datasets, and iteratively revise their analyses (Yao et al., 2023; Shinn et al., 2023; Wu et al., 2023). This development raises a more fundamental question than whether an LLM can generate scientifically plausible text: Can an AI agent reconstruct the complete reasoning process required to obtain a published scientific result from the information available to it?

Direct evaluation of AI systems for scientific discovery is challenging because novel scientific claims often lack an immediate ground truth. Scientific reproduction provides a complementary, controlled setting in which the target result is already known, while the agent must independently reconstruct the data, methodological, and computational steps required to obtain it (Peng, 2011; Stodden et al., 2014; Baker, 2016).

However, reproduction introduces a critical challenge for AI evaluation. Scientific papers are not executable specifications; rather, they describe results and methods for human readers who possess substantial background knowledge, disciplinary conventions, and practical experience. Consequently, a paper may explicitly state a parameter while leaving implicit the rationale for applying it to a particular dataset or downstream calculation. Multiple scientifically plausible choices often remain compatible with a single published description. For a human expert, such omissions are largely transparent because domain knowledge supplies the missing reasoning. For an AI agent, conversely, the same omission can generate multiple executable interpretations with distinct quantitative consequences. Consequently, a reproduction failure is inherently ambiguous: it may stem from an agentic error, source underspecification, or the failure of the model to correctly connect relevant information within the document.

This distinction motivates our study. Rather than evaluating scientific agents solely on their ability to reproduce a final numerical value, we investigate whether they can reconstruct the methodological dependencies linking scientific observations, assumptions, parameters, computational procedures, and conclusions. We use astronomy as a real-world testbed because astronomical research provides a rich combination of publicly accessible datasets, computational analyses, quantitative models, and heterogeneous methodological choices. We construct an experimental corpus of fourteen published astronomy studies (one in-depth ApJ case study plus a batch of thirteen papers published in Nature) and develop a multi-agent reproduction workflow that separates execution from verification while explicitly tracking intermediate artifacts and uncertainty.

## 2 Evaluation Framework

Unlike a mathematical proof, a software test, or a standardized benchmark, a genuinely new scientific hypothesis may take months, years, or even decades to validate experimentally. Even when an AI system produces a logically coherent and superficially plausible conclusion, contemporaneous validation is frequently impossible—leaving us unable to determine whether the output represents an original breakthrough, a compelling conjecture, or a fundamental error. This introduces a distinctive methodological bottleneck for AI-enabled science. Without immediate ground truth, establishing whether Model A outperforms Model B in scientific exploration becomes intractable, which in turn impedes iterative improvement. Resolving this foundational challenge requires building an experimental framework that renders scientific reasoning by AI observable, measurable, and assessable.

To operationalize this, we reframed the objective: rather than tasking AI with generating an entirely novel discovery from scratch, we evaluated whether an AI system could trace a scientific inquiry end-to-end—from formulating a question and sourcing data to understanding methods, executing analyses, interpreting outcomes, and recovering published conclusions. This reproduction protocol converts a reasoning process that traditionally lacks immediate feedback into a verifiable and comparative pipeline. To isolate core reasoning capabilities from confounding variables, the experimental design was structured across three controllable dimensions: the foundation model, the agent workflow, and the target publication.

Cross-model comparisons were systematically employed to mitigate architectural biases inherent to any single foundation model. Because raw generative models cannot autonomously execute end-to-end scientific workflows, we deployed a multi-agent architecture that operationalizes reasoning and coding into concrete actions, including literature searches, data acquisition, tool invocation, execution, and output validation (Boiko, MacKnight & Gomes, 2023). This multi-stage workflow transforms reproduction into an observable pipeline of inquiry, analysis, and validation, complete with explicit artifact generation and iterative discussion stages.

To counteract self-reinforcing evaluation biases in which agents excessively validate their own outputs, the pipeline explicitly decouples execution from adjudication. Critical milestones incorporate cross-model audits, preservation of intermediate artifacts, provenance tracking for parameters, execution auditing, and mandatory reporting of uncertainty and failure modes. As infrastructure-level errors were systematically isolated and resolved, our focus shifted from the limitations of the problem-solving systems to the intrinsic information constraints supplied by the source material itself.

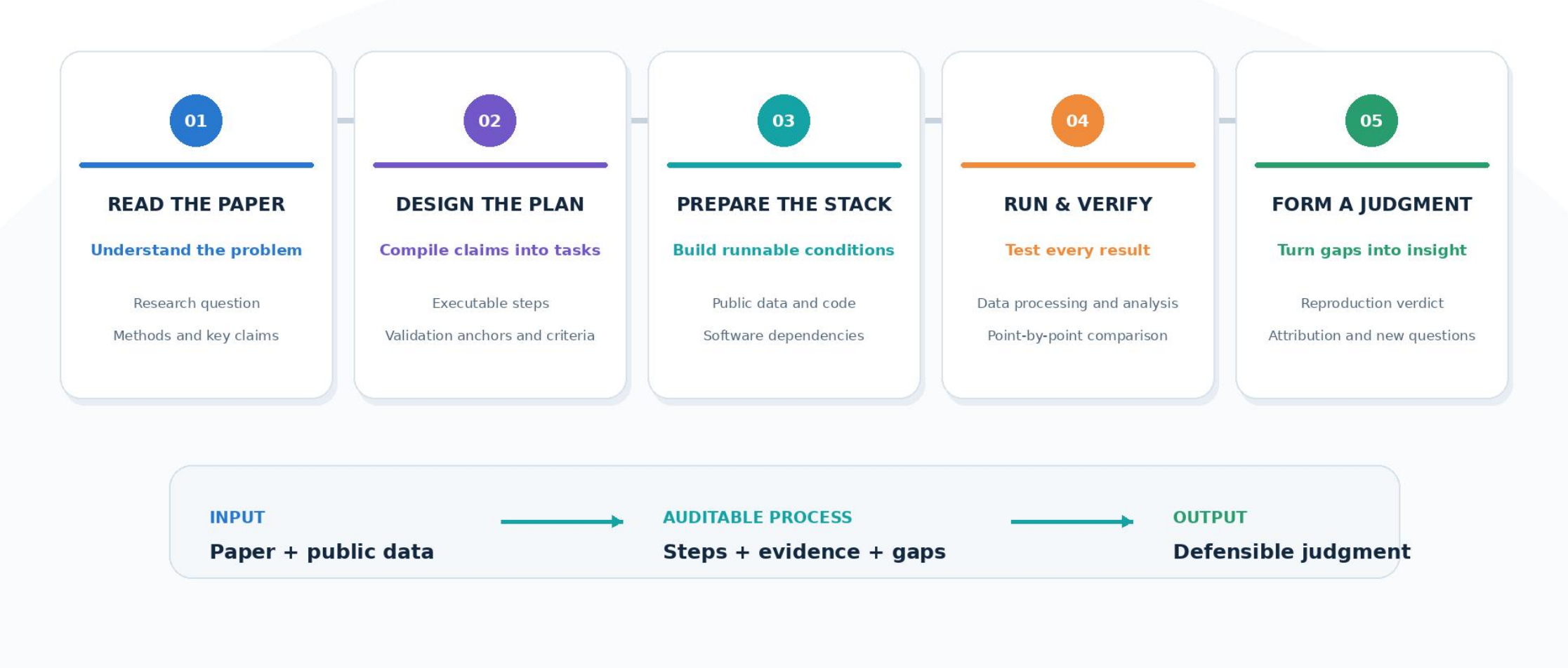

*Figure 1. Overall workflow for paper reproduction. The agent does not merely summarize the paper; it retraces the chain of evidence from inputs (paper and public data) through an auditable process (steps, evidence, and gaps) to an output in the form of a defensible judgment.*

# 3 Methods and Experimental Setup

This section specifies the components needed to reproduce our evaluation. Exact model identifiers, run logs, prompts, and per-path configurations are released with the code (see Data and Code Availability); where a value is finalized at submission time, it is marked here and in the reproducibility package.

## 3.1 Foundation models and configuration

Several frontier, instruction-following LLMs were used as the reasoning engines. Every analysis was performed with at least two independent foundation models so that no single conclusion rests on one model architecture. The exact model identifiers and versions, access dates, decoding temperature, context-window settings, random seeds (where supported), and the number of independent runs per paper are fixed at submission and recorded in the reproducibility package [to be completed at submission: model names/versions, temperature, top-p, context length, seeds, number of runs, run dates]. No model was given access to the published numerical target during path generation in the controlled case study.

## 3.2 Agent architecture and roles

The workflow was orchestrated with Hermes, an open-source multi-agent framework (Nous Research, 2026; https://github.com/NousResearch/hermes-agent), which schedules specialized roles and provides them with tools and a persistent filesystem. Five roles were used: a reader, which parses the target paper and accessible supplementary material into structured notes; a planner, which decomposes the reproduction into an executable task graph; a coder, which writes and executes the analysis; a verifier, which independently re-derives and checks outputs against the paper; and an adjudicator, which resolves execution–verification disagreements through cross-model audit rather than self-confirmation.

## 3.3 Toolchain and execution environment

Agents had access to web and scholarly literature retrieval, public astronomical data archives (including the Gaia archive and, where relevant, MAST, TESS, and survey releases), a persistent filesystem, and a sandboxed Python execution environment with pinned package versions and shell access. Each target paper was processed in an isolated working directory and software environment. Every downloaded dataset, generated script, intermediate output, and parameter decision was logged with provenance. The full prompts, role definitions, package manifest, configurations, and intermediate artifacts are released with the paper [to be completed at submission: GitHub/Zenodo DOI and environment lockfile].

## 3.4 Audit protocol for the thirteen Nature papers

Each of the thirteen Nature papers was independently processed in two rounds driven by different foundation models; the adjudicator reconciled the two records, and a human domain expert reviewed every retained item. A paper was classified as lacking a uniquely specified reproduction path when at least one methodological, configurational, or interpretational ambiguity prevented an independent agent from proceeding from the public materials to the reported numerical result without making a choice that the paper did not uniquely determine. Issue categories are not mutually exclusive (a single paper can contribute to several), so the denominator for all category counts is the number of papers (n = 13), and category counts can sum to more than thirteen. Items whose status could not be resolved—for example, because they appeared to arise from text extraction rather than from the source—were recorded as unverified rather than counted as confirmed.

### 3.5 Controlled design of the twelve analysis paths

For the in-depth case study, the degrees of freedom identified in the published method were (i) the stellar-sample definition, with three alternatives; (ii) the sky mask, with two alternatives; and (iii) the treatment of the Gaia parallax zero-point, with two alternatives. Their full factorial combination, $3 \times 2 \times 2$, defines twelve predefined, independently executable paths. The twelve paths constitute a sensitivity analysis, not an attempt to reverse-engineer the authors' choices: the published value (≈2.7 kpc) was not used as an optimization target, a selection criterion, or a stopping condition during path generation, and the single path that matched the published value was identified only after all twelve predefined paths had been executed.

# 4 Experimental Corpus and Case Studies

The study began with one paper from The Astrophysical Journal (ApJ), which surfaced the initial problem, and then expanded to thirteen astronomy papers published in Nature to test whether the same pattern would recur (Figure 2). Papers were drawn from leading journals not because reproducing them was an end in itself, but because they provide peer-reviewed scientific questions, methodological descriptions, and quantitative conclusions that serve as fixed reference points. Three practical eligibility criteria were applied: (1) the underlying data were publicly accessible; (2) the report presented sufficiently concrete material to define a meaningful end-to-end reproduction task from the available public material; and (3) the headline conclusion was broadly accepted by the field, so that a reproduction discrepancy could not simply be attributed to contested science. These criteria concern feasibility and the existence of a stable reference, not whether a paper could be reproduced; they therefore do not pre-select papers by result.

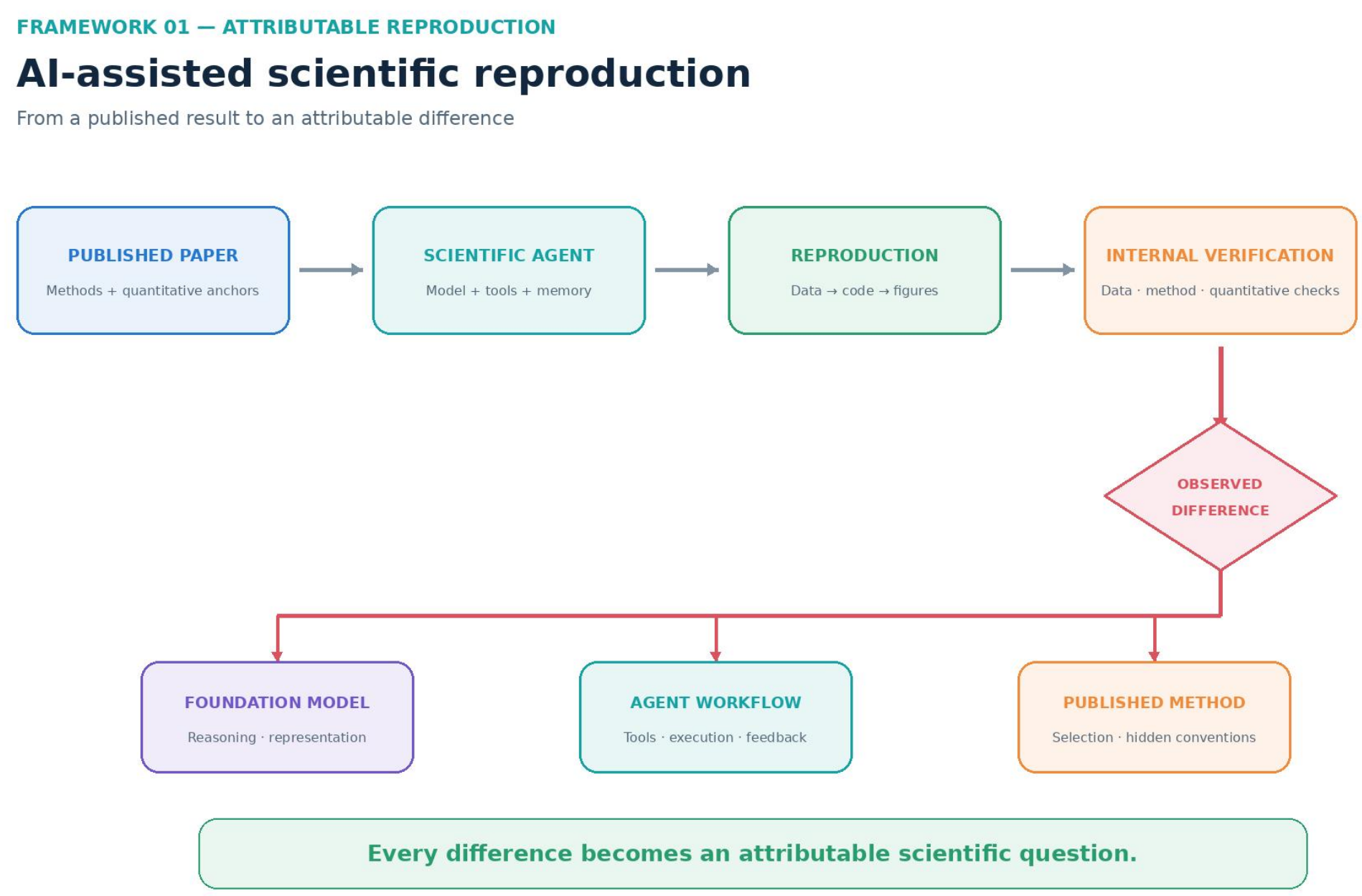

*Figure 2. Reproduction and attribution in AI-assisted scientific research. The corpus comprises one initial ApJ case study and a batch of thirteen astronomy papers published in Nature; the latter supplies the principal corpus-level evidence.*

## 4.1 The initial ApJ case: identifying the problem

The first paper was "The Poor Old Heart of the Milky Way," published by Rix et al. in The Astrophysical Journal in 2022 (Rix et al., 2022). Using the Hermes framework, we connected several large language models and ran multiple independent reproduction attempts. Every run recovered the paper's central qualitative result: metal-poor stars are concentrated toward the Galactic centre, and the two stellar populations show clear differences in chemical composition and orbital properties. At the quantitative level, however, repeated runs failed to align with the values reported in the paper. The agents began tracing the discrepancies backward through the workflow (Figure 3).

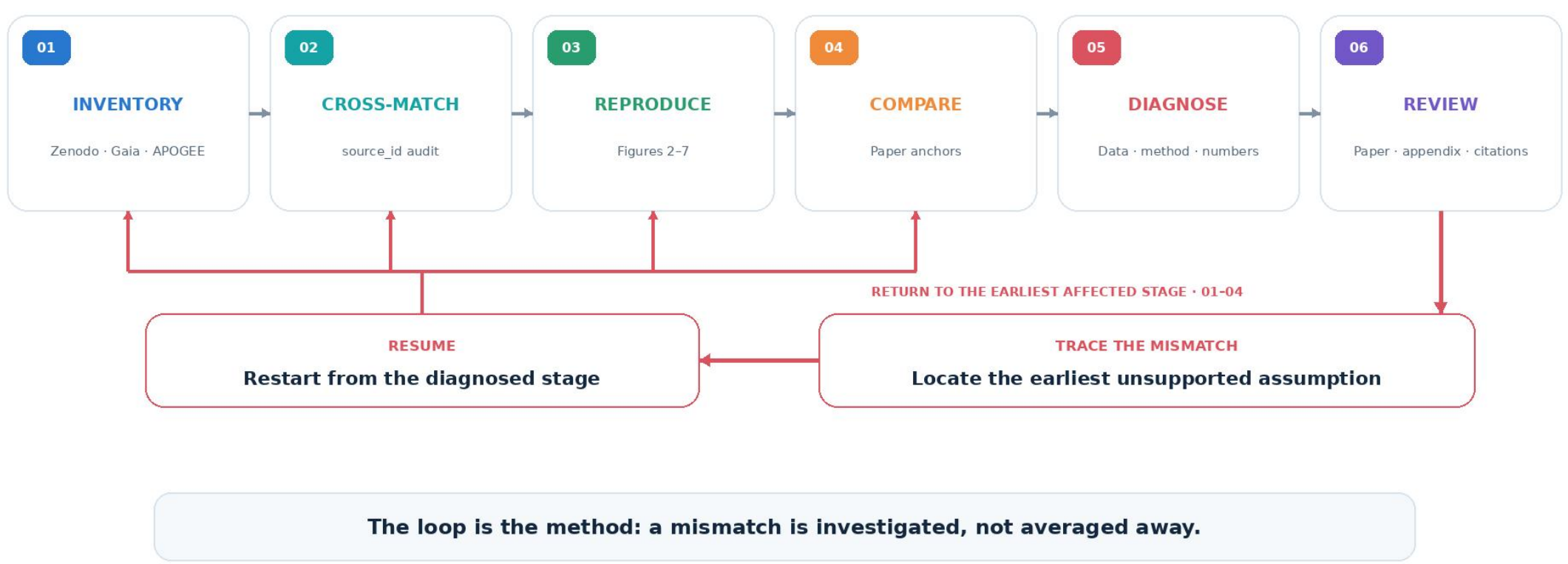

*Figure 3. The workflow used for the Rix et al. (2022) reproduction.*

The first anomaly appeared in sample selection. Following the method stated in the paper produced nearly 30,000 stars, whereas the published analysis used approximately 18,000–20,000. On the basis of the paper text and currently available public data alone, we could not reconstruct the same counting convention. The discrepancy could reflect an unstated subsample definition, additional quality cuts, deduplication rules, or differences in data versions—details that were not fully and unambiguously specified in the article (Figure 4).

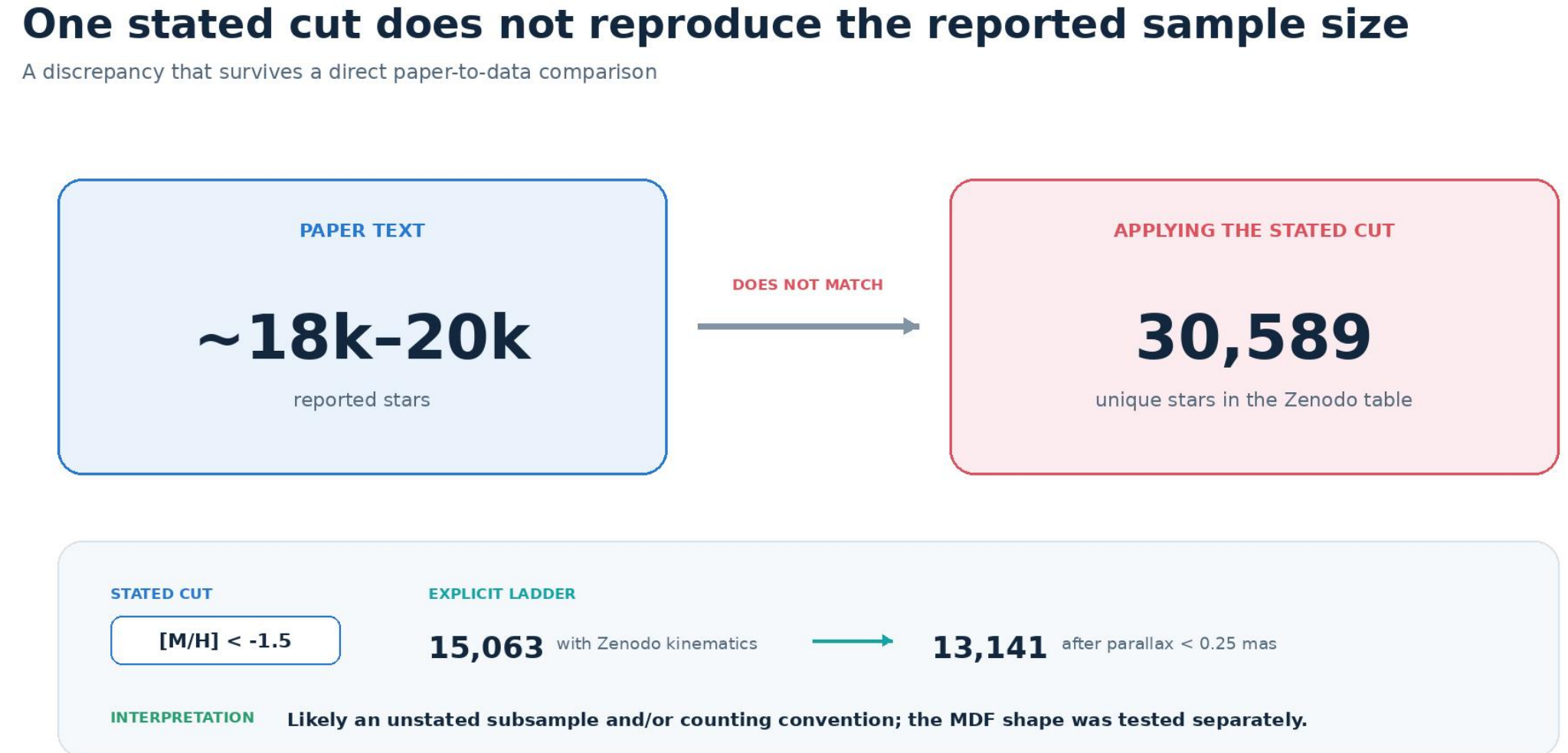

*Figure 4. Difference between the paper's reported sample size and the recoverable selection conventions: one stated cut does not reproduce the reported sample size.*

A second anomaly involved the paper's central spatial-scale parameter, the radial dispersion of the metal-poor population, denoted $\sigma_{RGC}$ (see Figure 5). The paper reported a value of approximately 2.7 kpc. The initial reproduction recovered the qualitative trend, but the public method did not uniquely determine the executable choices governing sample definition, sky masking, and parallax zero-point treatment; accordingly, it did not uniquely determine the 2.7 kpc value either. At that stage, we recorded the mismatch as an unresolved discrepancy. We did not choose parameters according to how closely they reproduced the published number.

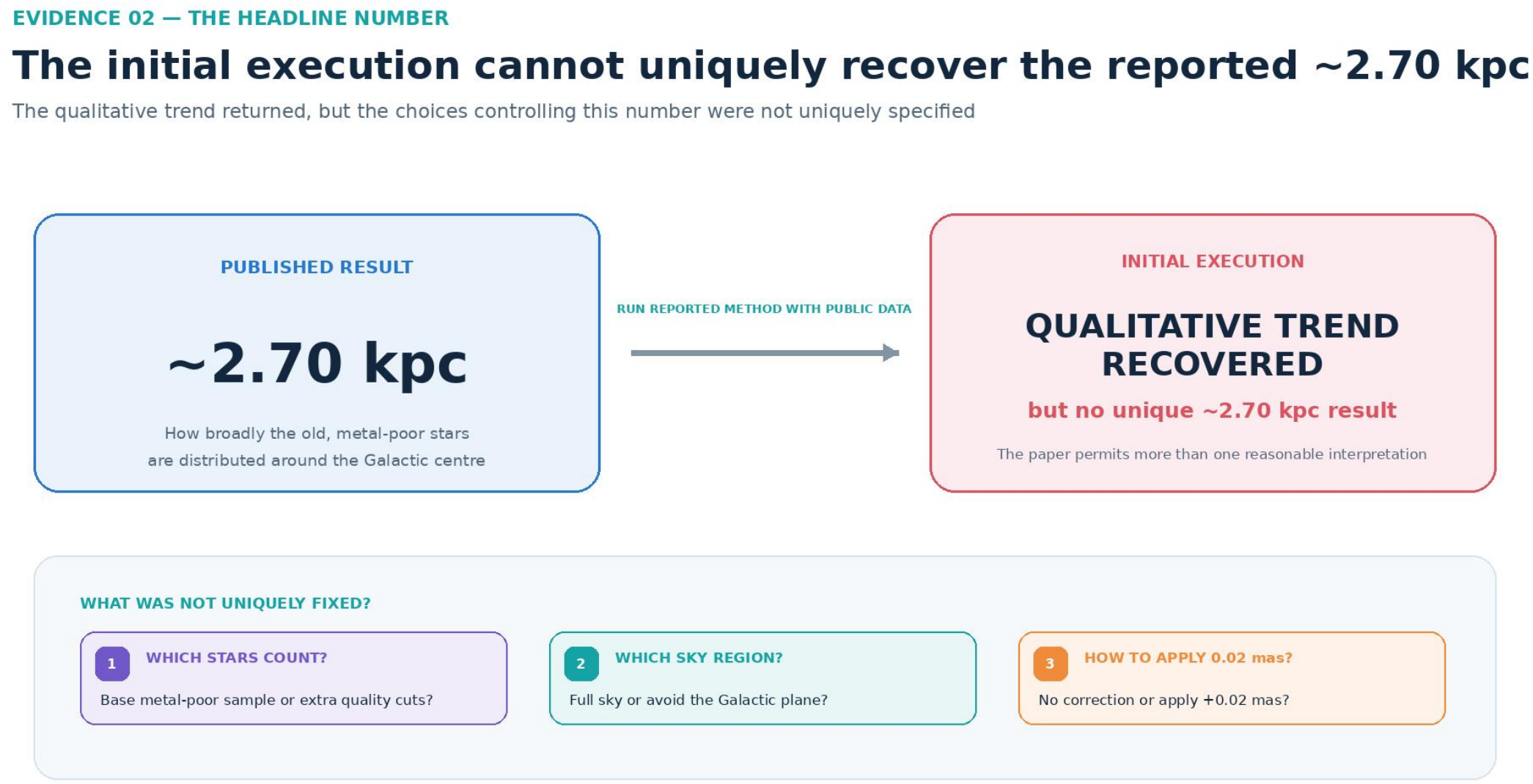

*Figure 5. Quantitative reproduction discrepancy for the spatial scale $\sigma_{RGC}$.*

At the time, we could not locate the source of this first reproduction failure precisely. It might have reflected genuinely missing practical information in the paper, and thus a potentially important scientific clue. But it could just as easily have arisen from our own lack of domain expertise, a hidden defect in the agent pipeline, or a failure somewhere in the model–tool chain. We therefore set a new goal: reproduce a small group of papers from the same broad field while systematically refining the workflow and reducing interference from the pipeline, agents, models, and human operation.

## 4.2 Expanding to thirteen Nature papers

When the first paper could not be reproduced exactly, the agent system was our primary suspect. As we refined the pipeline, however, our attention shifted from the system solving the problem to the information supplied by the problem itself. We selected thirteen astrophysics papers published in Nature and ran a batch reproduction experiment. Across the thirteen papers, eleven contained at least one issue that prevented a uniquely specified reproduction path. Only two provided methodological descriptions that were sufficiently complete, within the scope of our tasks, to support an independent path from data to the target numerical result (Figure 6).

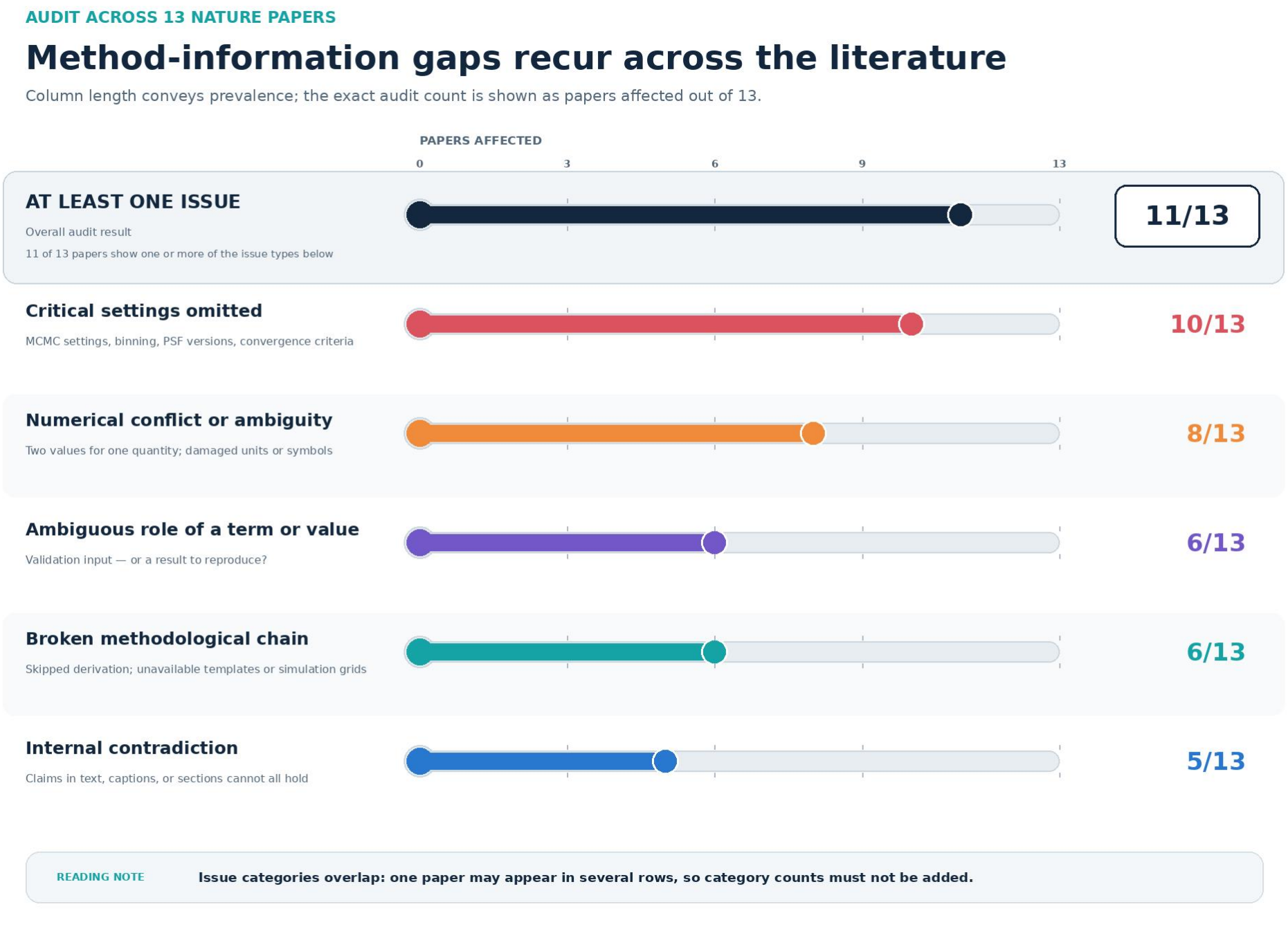

*Figure 6. Audit of methodological information in the thirteen-paper Nature sample. Eleven of thirteen papers contained at least one issue that prevented a uniquely specified reproduction path.*

This changed how we thought about reproduction. Initially, we had hoped to use it as an indirect test of whether AI could perform scientific discovery. The batch experiment suggested that reproduction might be more than a way to verify existing discoveries: it could also provide an entry point for finding new knowledge. When an AI system retraces the path from data to scientific conclusion, research judgements that were omitted or treated as obvious may become visible again. A discrepancy between a reproduced result and a published result is not necessarily an error to be eliminated. Once verified and explained, it may reveal a relationship among the source data, the analytical method, and the scientific conclusion that has not been made fully explicit. In that sense, reproduction can move from validating existing knowledge toward discovering knowledge hidden within existing research.

## 4.3 Returning to the first paper: a controlled sensitivity analysis

With this idea in mind, we returned to the first paper using the revised reproduction system. This time, we stopped trying to tune the result toward the paper's 2.7 kpc value. Instead, we asked a different question: if the paper does not uniquely specify an analysis path, how much can scientifically plausible choices change the conclusion? We turned this question into a controlled experiment (Figure 7). One of the central results in Rix et al. is the spatial extent of a metal-poor stellar population around the Galactic centre, reported as approximately 2.7 kpc. Several choices were not uniquely determined by the methodological description: which stars should be included, whether a systematic parallax offset should be corrected, and whether regions close to the Galactic plane should be excluded. We enumerated reasonable options for these three dimensions and generated twelve possible analysis paths. These paths were not twelve guesses about what the authors had actually done; they were a systematic exploration of the consequences of choices that the paper did not reduce to a single executable specification.

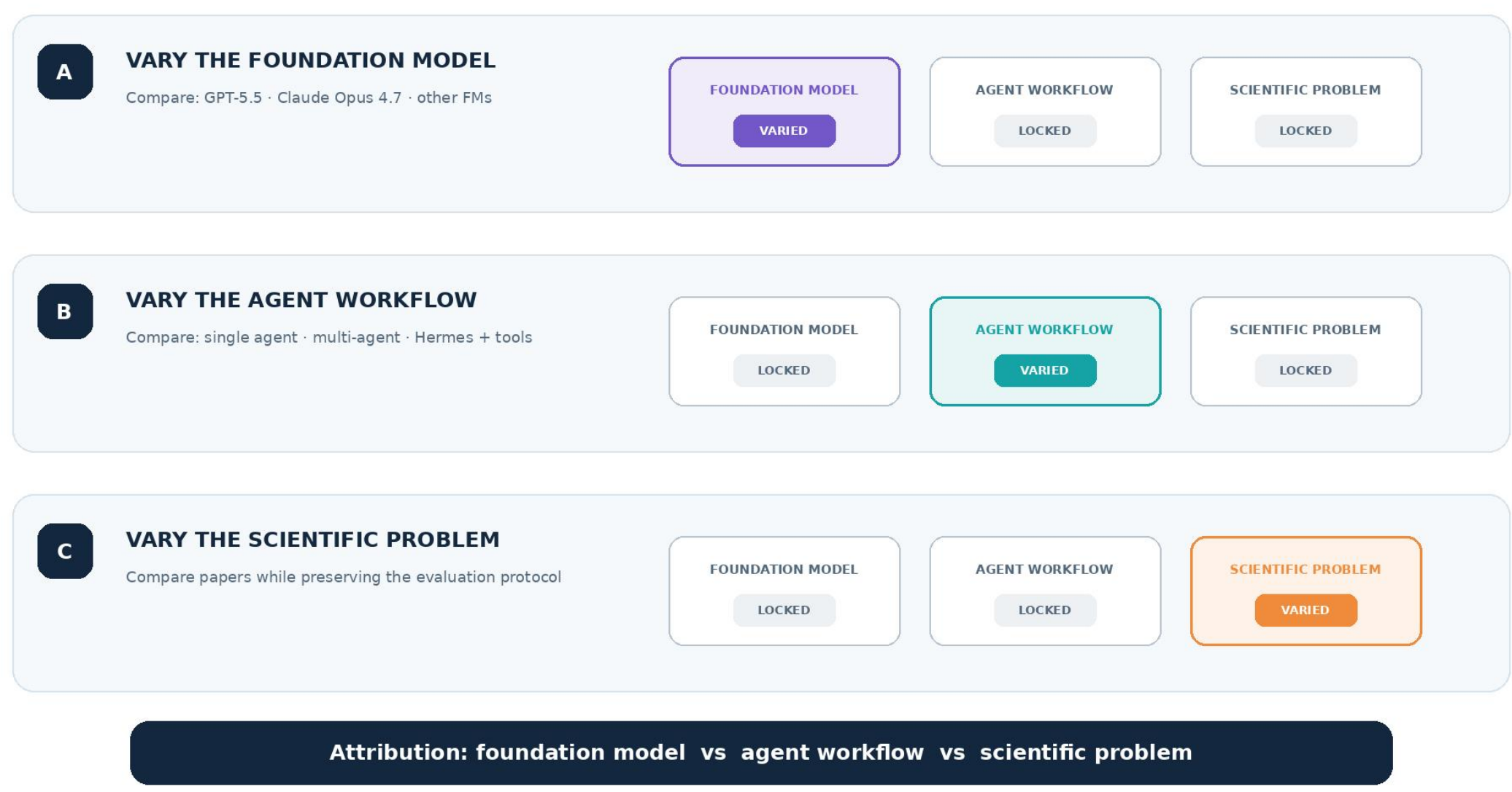

*Figure 7. Attribution logic for three families of controlled comparisons: change one variable while holding the other two fixed.*

All twelve paths converged and executed successfully, producing spatial scales from 2.16 to 3.53 kpc. One path—a particular sample definition, a parallax zero-point correction, and the full-sky region—produced approximately 2.70 kpc, closely matching the reported result (Figure 8).

To foreclose any reverse-engineering interpretation, we state explicitly: the published value was not used as an optimization target, selection criterion, or stopping condition during the generation of the 12 paths. The matching path was identified only after all 12 predefined paths had been executed. Path labels, parameter grids, and stopping rules were fixed before any path was run, and no path was added, removed, or re-weighted after the published number became visible to the analysis team.

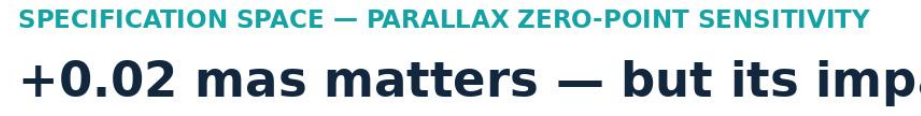

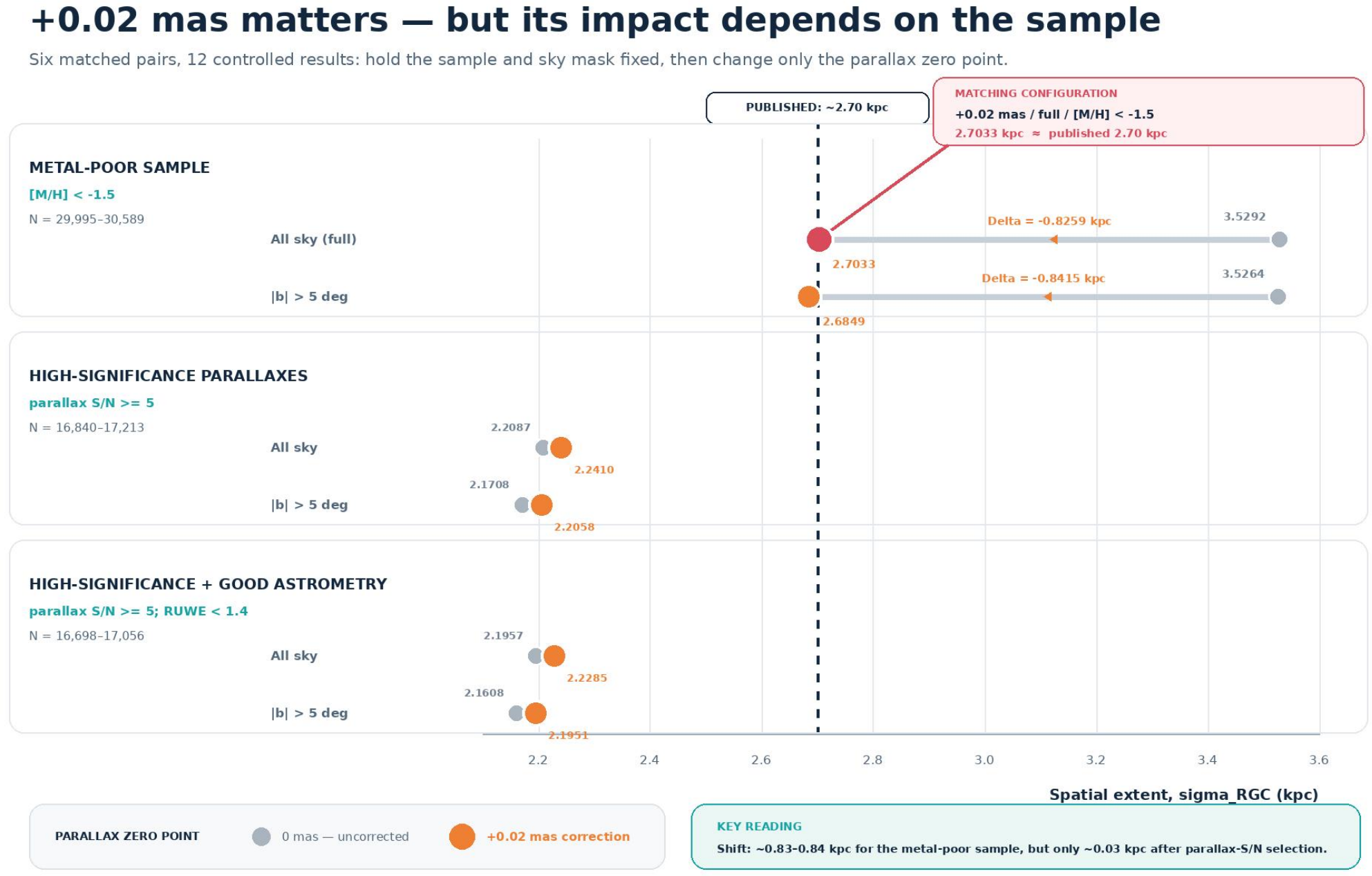

*Figure 8. Sensitivity of the twelve analysis paths to parallax zero-point treatment (3 sample definitions × 2 sky masks × 2 zero-point treatments).*

Our primary concern, however, was not which analytical path came closest to the 2.7 kpc benchmark, but why—given that all twelve paths executed successfully—any single path should be preferred over the remaining eleven. At this juncture, the AI surfaced a connection it had previously failed to establish. The relevant information was not absent from the text: Rix et al. had already reported a value of +0.02 mas and cited Lindegren et al. (2021) on the Gaia EDR3 parallax zero-point. Rather than the parameter itself, what had been compressed was the underlying methodological judgment—why a constant approximation remained appropriate for this specific sample, under what conditions that approximation held, and when it required re-estimation for subsequent data releases or alternative samples (Lindegren et al., 2021; ESA, 2021). This crucial rationale did not reappear as an explicit step within the quantitative methodology. To an experienced astronomer the inference may have been intuitive, because the target distance scale, Gaia parallaxes, and the known parallax zero-point bias naturally link the correction discussion to the downstream fit. The system, however, did not make this connection autonomously. Although it successfully parsed "+0.02 mas" and executed the subsequent fitting procedure, it did not infer that a value introduced earlier in the text needed to govern the quantitative analysis, or that omitting it would materially alter the outcome. Only after the twelve-path experiment rendered this effect quantitatively visible did the system bridge information distributed across disparate sections of the paper. The salient finding is therefore not that the parameter was missing, but that its causal relevance was not recognized. While modern AI systems can systematically explore a space of analytical choices and execute alternatives in parallel, The evaluated agents continued to struggle with recognizing why one empirical fact is causally relevant to another in the absence of explicit guidance to investigate that relationship.

# 5 Related Work

**LLM agents for science.** Transformer language models and instruction-following systems established the underlying capabilities used throughout this work (Vaswani et al., 2017; Brown et al., 2020; OpenAI, 2023). Building on tool use, retrieval, and self-reflection, language-model agents now plan and execute multi-step tasks (Yao et al., 2023; Shinn et al., 2023; Wu et al., 2023), including in scientific domains, where LLM-driven systems have been used to design and run chemical experiments and to operate laboratory instrumentation (Boiko, MacKnight & Gomes, 2023). Our work treats such an agent as the subject of evaluation rather than as a black-box performer.

**AI for scientific discovery.** A parallel line of work pursues autonomous "AI scientists" that generate hypotheses, run analyses, and draft studies end-to-end (Lu et al., 2024; Si et al., 2025). Demonstrations of autonomous discovery are difficult to evaluate rigorously because genuine novelty has no immediate ground truth. Reproduction supplies a controlled lens: the target is known, so the quality of the reconstructed path—not only the final number—becomes measurable.

**Reproducibility and computational reproducibility.** Surveys across the sciences find that many published results are difficult or impossible to reproduce (Baker, 2016), and methodological work addresses dependency capture, provenance, and executable environments (Peng, 2011; Stodden et al., 2014). We use reproducibility for a distinct purpose: rather than treating reproduction only as a reliability target, we use it as a diagnostic instrument that reveals which steps a publication leaves implicit.

**Implicit and tacit knowledge.** Philosophical and sociological accounts emphasize that expert competence contains a tacit component that is not fully expressible in text (Polanyi, 1966; Collins, 2010). Recent critiques of AI-assisted science argue that publications compress methodological judgment. Our contribution is operational: we turn the presence and recovery of such implicit knowledge into an observable variable by measuring whether an end-to-end reproduction agent re-establishes the missing methodological dependencies.

# 6 Results

The batch reproduction of thirteen papers and the deeper investigation of Rix et al. point to the same conclusion: within this corpus, incomplete specification of the implicit knowledge required for quantitative analysis is a recurring pattern rather than an isolated case. Across the thirteen-paper Nature sample, eleven studies contained at least one methodological, configurational, or interpretational ambiguity that prevented a uniquely specified reproduction path within the scope of our experiments (Figure 6). In the controlled case study, twelve independently executable paths yielded spatial scales between 2.16 and 3.53 kpc, with one path recovering the published value of approximately 2.70 kpc (Figure 8). The experiment shows that matching a published number does not by itself establish that the underlying reasoning path has been reconstructed.

Our experiments reveal a specific limitation of the scientific agents evaluated in this study: while the LLMs could retrieve explicit information and execute well-defined computational procedures, they were unreliable at recognizing when information distributed across a scientific document is causally relevant to a downstream analysis. The hardest part of paper reproduction, then, may not be finding a parameter, but understanding why that parameter belongs to this dataset, this sample, and this scientific question. This does not mean that the papers' conclusions are unreliable. A paper must prioritize novelty; many choices involving parameters, data filtering, calibration, and uncertainty handling are not themselves novel, yet they are decisions that cannot be made incorrectly without consequences. For domain scientists, such judgements are often acquired through training, convention, and experience. For AI, they do not automatically emerge even when the paper, data, and code are all available. Auditing can prevent an AI system from forcing missing pieces into place simply to obtain an answer, but it cannot substitute for domain experience.

# 7 Limitations

Several limitations bound the scope of our claims. First, the corpus is small and staged by design: one in-depth ApJ case study plus a batch of thirteen Nature papers, rather than a single uniform benchmark of fourteen. Second, the corpus is confined to astronomy, a data-rich field with unusually open archives; the frequency and character of implicit dependencies may differ in other disciplines. Third, although audit categories were produced by two model-driven rounds, reconciled by an adjudicator, and confirmed by human review, their assignment remains partly interpretive and could be coded differently by another team. Fourth, LLM capabilities and versions evolve rapidly, so absolute failure rates are time-dependent and may change with later models; we therefore report a mechanism and an evaluation method rather than a permanent benchmark number. Fifth, reproduction is sensitive to data releases and to the completeness of publicly accessible material, both of which can change over time. Finally, the twelve-path sensitivity analysis concerns a single quantity in a single paper; while the mechanism it exposes—present-but-unlinked causal information—is suggestive, broader generalization requires larger and more diverse corpora.

# 8 Discussion and Conclusion

At VivaTech in 2023, Yann LeCun argued that much of human knowledge is non-linguistic and that this experiential component has therefore not been captured by current AI systems. The same concern applies to our experiments on scientific capability. Papers preserve what scientists discovered, but they do not always preserve the full process by which scientists knew that a particular choice was the right one. This may be an easily overlooked risk when AI attempts to conduct scientific discovery independently.

We argue that end-to-end scientific reproduction can serve not only as a reproducibility test but also as an evaluation framework for implicit scientific knowledge and scientific reasoning in AI systems. The controlled Rix case illustrates why this framing is useful: a predefined sensitivity analysis, in which the published target was never used for tuning, shows that a numerical match is neither necessary evidence of correct reasoning nor a stable target to optimize toward, whereas the failure to connect a correction already present in the paper to the downstream fit is directly observable. Our next goal, therefore, is not simply to reproduce more papers, but to enable models, through large-scale reproduction, to actively reconstruct knowledge embedded in scientific practice but compressed in publication. As the scale grows from dozens of papers to hundreds or thousands, recurring parameter conflicts, methodological variations, and failure modes may begin to reveal patterns that transfer across papers and disciplines. We do not claim that reproduction has produced new scientific discoveries here; rather, we show that it can serve as a disciplined entry point through which such implicit knowledge—and, potentially, new scientific questions—may come to light.

# Appendix A — Implicit-Knowledge Clues Across Thirteen Astronomy Papers Published in Nature

All thirteen papers have DOIs beginning with 10.1038/s41586-, the identifier used by the main weekly journal Nature; journals elsewhere in the Nature portfolio use other identifiers, such as s41550 for Nature Astronomy, s41567 for Nature Physics, and s41467 for Nature Communications. We therefore refer to the batch throughout as “thirteen astronomy papers published in Nature,” not as Nature Astronomy papers.

Status of the audit observations. The clues listed below are audit observations generated during the reproduction experiments. They must not be interpreted as claims that the corresponding papers contain errors. Some observations arise from text extraction, from incomplete access to supplementary material, or from unresolved methodological ambiguity, and they require independent verification against the original articles. Each item is the reconciled product of two model-driven rounds plus human review; items that could not be confirmed are explicitly marked. A complete, machine-readable audit table, the per-path configurations for the twelve-path experiment, and all intermediate artifacts are released as supplementary material (see Data and Code Availability); the summary statistics reported in the main text are those in Figure 6.

*Reading guide. The columns report (left to right) the paper number, title, DOI and citation, topic, the study’s main finding, the implicit-knowledge clues identified during reproduction, and the mining approach used to surface them. Individual clues are separated by line breaks and do not represent independent error counts.*

| No. | Paper title | DOI / citation | Topic | Main finding | Implicit-knowledge clues | Mining approach |
| --- | --- | --- | --- | --- | --- | --- |

| No. | Paper title | DOI / citation | Topic | Main finding | Implicit-knowledge clues | Mining approach |
|---|---|---|---|---|---|---|
| 01 | Detection of anisotropic cosmic structures on a gigaparsec scale | 10.1038/s41586-026-10702-5<br>Nature, 2026 | Cosmology · Large-scale structure | Using the DESI galaxy survey and a parameter-free statistic, the study detects anisotropy in the galaxy distribution on billion-light-year scales beyond that expected in simulated universes, challenging the cosmological principle. | – Three different counts are given for the simulation sample: 330 in the main text, 1,000 in Fig. 4a, and 200 in Fig. 4b—more than one number for the same test.<br>– The paper does not clearly identify what is meant by a “representative BGS subsample.”<br>– The number of angular bins is not given, nor are the radial-bin boundaries or partitioning rule.<br>– A formal record of the same issue as the first clue: the sample-count convention is inconsistent with the significance threshold.<br>– The normalization coefficient in the angular-variance formula (Eq. 3) is corrupted in the extracted text, so its exact form cannot be read reliably.<br>– The simulated particle count is typographically ambiguous: should “1,0243” be read as $1024^3$ or something else? | Cross-check conventions for simulation counts, significance thresholds, and related quantities.cross the two rounds of experimental records. |

| No. | Paper title | DOI / citation | Topic | Main finding | Implicit-knowledge clues | Mining approach |
|---|---|---|---|---|---|---|
| 02 | GW250114 reveals signatures of post-merger black-hole horizon | 10.1038/s41586-026-10696-0<br>Nature, 2026 | Gravitational waves · Black holes | GW250114 reveals signatures of post-merger black-hole-horizon behaviour and tests Einsteinian gravity. | – The units of the damping rate are not specified.<br>– Two interpretations assign different roles to the same “300 simulated backgrounds” setup.<br>– The paper provides only a theoretical citation and does not state where the NRSur7dq4 waveform was obtained.<br>– It is unclear which quantity the "remnant map" parameter calibrates; this does not align with the angle emphasized in the main text.<br>– The paper says the result is "fully consistent" with Kerr theory without providing a consistency metric, while also stating that ωHωH need not match exactly—the two formulations conflict.<br>– Details of the exact analytical model depend on the Supplementary Information, but the two interpretations disagree on the degree of that dependence. | Merge the two rounds of experimental records and place the two theoretically contradictory formulations side by side. |

| No. | Paper title | DOI / citation | Topic | Main finding | Implicit-knowledge clues | Mining approach |
|---|---|---|---|---|---|---|
| 03 | A direct black-hole mass measurement in a little red dot at high redshift | 10.1038/s41586-026-10579-4<br>Nature, 2026 | JWST · Black-hole measurement | JWST directly measures the mass of a black hole in a high-redshift "little red dot" galaxy. | – The same ≈10 km/s value appears with two meanings: velocity gradient versus projected velocity per bin.<br>– An object reconstructed by only one interpretation has no independent corroboration from the other.<br>– The central value is readable, but the uncertainty values are unreliable in the visible text.<br>– For the 0.05 arcsec data cube, one interpretation treats the value as a validation input and the other as an execution parameter; its role is unclear.<br>– The two interpretations assign opposite labels to whether the reported stellar mass should be treated as a "reference answer" to prevent leakage.<br>– The mass and inclination likewise receive conflicting reference-answer labels. | Merge the two rounds of experimental records and isolate disagreements over "reference answer versus validation input," together with the 1 dex mass discrepancy. |

| No. | Paper title | DOI / citation | Topic | Main finding | Implicit-knowledge clues | Mining approach |
|---|---|---|---|---|---|---|
| 04 | An ultra-faint, chemically primitive galaxy forming in the reionization era | 10.1038/s41586-026-10374-1<br>Nature, 2026 | JWST · Early galaxies | The study discovers an extremely faint, metal-poor dwarf galaxy that formed during the epoch of cosmic reionization. | – The two interpretations disagree on the applicable scope of method M07.<br>– They also disagree on how strongly the "hard spectrum" interpretation is supported.<br>– The complete methodological chain for the dynamical mass is missing. | Merge the two rounds of experimental records and enumerate three sets of methodological-semantic disagreements: the role of the oxygen-line ratio, the hard-spectrum interpretation, and the dynamical-mass chain. |
| 05 | Satellite megaconstellations will threaten space-based astronomy | 10.1038/s41586-025-09759-5<br>Nature, 2025 | Space diplomacy · Satellite constellations | Satellite megaconstellations such as Starlink threaten ground- and space-based astronomical observations. | – The paper is internally inconsistent: the SPHEREx orbital altitude is given as 650 km in one place but 700 km in the overview.<br>– The Hubble exposure-time bounds are also reconstructed values; the asymmetric bounds still require verification against the original manuscript. | Merge the two rounds of experimental records and recover, item by item, the values corrupted during text extraction, including orbital altitude, distance bounds, and exposure-time bounds. |

| No. | Paper title | DOI / citation | Topic | Main finding | Implicit-knowledge clues | Mining approach |
|---|---|---|---|---|---|---|
| 06 | Witnessing the onset of reionization through Lyman-α emission at redshift 13 | 10.1038/s41586-025-08779-5<br>Nature, 2025 | JWST · Cosmic reionization | Lyman-α emission at redshift 13 provides a view of the onset of cosmic reionization. | – The two interpretations disagree on how the systemic redshift was determined.<br>– Neither interpretation can recover the EDT3 prior distribution or most posterior-table values; the paper does not specify the implementation of σ-clipping, covariance construction, or path-loss correction.<br>– The ForcePho photometric and morphological-fitting configuration is incomplete: neither the version nor the PSF is provided. | Merge the two rounds of experimental records and isolate three sets of parameters lacking a uniquely specified source: priors, σ-clipping, and photometric configuration. |

| No. | Paper title | DOI / citation | Topic | Main finding | Implicit-knowledge clues | Mining approach |
|---|---|---|---|---|---|---|
| 07 | Spectroscopic confirmation of two luminous galaxies at a redshift of 14 | 10.1038/s41586-024-07860-9<br>Nature, 2024 | JWST · High-redshift galaxies | Spectroscopy confirms two luminous galaxies at redshift 14. | – The units of the slit-loss compensation polynomial are not specified for the z=14.0z=14.0 galaxy.<br>– The shared parameter U5 has conflicting semantics: the same name is used with different meanings.<br>– The paper provides no numerical reproduction tolerance; another interpretation uses comparison language but supplies no pass/fail criterion.<br>– The single exposures, per-exposure PSFs, software version, and configuration required for morphological fitting are not provided.<br>– One interpretation cites a source outside the permitted rules and requires verification.<br>– The sampler, likelihood implementation, and convergence criteria are missing, preventing numerical reproduction. | Merge the two rounds of experimental records and juxtapose four groups of missing configuration details: units, tolerances, priors, and PSFs. |

| No. | Paper title | DOI / citation | Topic | Main finding | Implicit-knowledge clues | Mining approach |
|---|---|---|---|---|---|---|
| 08 | Fast-moving stars around an intermediate-mass black hole in ω Centauri | 10.1038/s41586-024-07511-z<br>Nature, 2024 | Black holes · Globular clusters | Fast-moving stars in the ω Cen globular cluster indicate the presence of an intermediate-mass black hole. | – The two interpretations disagree on whether the robustness tests were extended to the surface-brightness model, mass-to-light ratio, and distance.<br>– The full MCMC configuration is not provided, yet one interpretation labels it “determinate.”<br>– The grid bounds and step size for the minimum-mass method are unspecified.<br>– The external simulation grid used for the N-body comparison is unavailable; neither interpretation provides sufficient numerical-sampling configuration to reproduce the posterior deterministically.<br>– Extended Data Table 1 contains missing cells, although both interpretations require those values for their calculations. | Merge the two rounds of experimental records and identify the three methods labelled "determinate" despite missing configuration details. |

| No. | Paper title | DOI / citation | Topic | Main finding | Implicit-knowledge clues | Mining approach |
|---|---|---|---|---|---|---|
| 09 | A resonant sextuplet of sub-Neptunes transiting the bright star HD 110067 | 10.1038/s41586-023-06692-3<br>Nature, 2023 | Exoplanets · Resonant chains | The six planets orbiting HD 110067 form a 3:2/4:3 period-resonance chain; two were recovered by reprocessing TESS data. | – The main text and figure caption conflict: the text states 120 s sampling, whereas the caption gives 2 minutes for Sector 23 and 20 seconds for Sector 49.<br>– The predicted period of the outer planet g is 54.7433 days, versus the measured value of 54.76992 days in Table 1. This has already been investigated in depth: the difference can be decomposed into the product of the detunings of adjacent orbital pairs.<br>– The ± ordering of the asymmetric uncertainties in Table 1 was reconstructed from corrupted text cells and still requires verification against the original table. | Place the two period values side by side and check them arithmetically: the 0.0266-day difference equals 133 times the error bar. Then, independently verify the transit times using locally available public TESS data, recovering agreement within 2–4 minutes, and compare the two experimental records to identify the contradiction in the captioned sampling cadence. |
| 10 | The solar dynamo begins near the surface | 10.1038/s41586-024-07315-1<br>Nature, 2024 | Solar physics | The solar magnetic-field dynamo begins near the surface rather than in the deep interior, as traditionally assumed. | – The two interpretations agree on the numerical rotation-profile equation, but the formula itself is problematic: the exact algebraic solution of the quasi-linear equation lacks a parameter. | Merge the two rounds of experimental records and confirm the parameter missing from the assembled formula. Repeat the extraction with a different OCR tool. |

| No. | Paper title | DOI / citation | Topic | Main finding | Implicit-knowledge clues | Mining approach |
|---|---|---|---|---|---|---|
| 11 | Orbital period change of Dimorphos due to the DART kinetic impact | 10.1038/s41586-023-05805-2<br>Nature, 2023 | Planetary science · DART mission | The study measures the change in Dimorphos's orbital period after the DART spacecraft impacted the binary-asteroid system. | – Two precision levels coexist for the pre-impact period: 11.92148 ± 0.00013 h (3σ) and a rounded abbreviated value.<br>– The two interpretations classify the period differently—as an "external input" or a "method output."<br>– The paper gives no reproduction tolerance. One interpretation proposes passing results "within 3σ"; one includes an initial radar estimate of −36 ± 15 minutes, while the other omits it entirely. | The formal experiment in this paper was not completed, so an earlier shadow-experiment record—which did complete the run—was used retrospectively. |

| No. | Paper title | DOI / citation | Topic | Main finding | Implicit-knowledge clues | Mining approach |
|---|---|---|---|---|---|---|
| 12 | Star formation near the Sun is driven by expansion of the Local Bubble | 10.1038/s41586-021-04286-5<br>Nature, 2021 | Milky Way · Star formation | Star formation near the Sun is driven by the expansion of the Local Bubble. | – The claimed "qualitative morphology" has no metric or threshold. The expansion model is also idealized, and the equation's exponent was inferred backward from a standard form in an external citation. Neither interpretation provides a numerical reproduction tolerance.<br>– The specific variant of the Milky Way-like gravitational potential and its integration settings are unspecified; external numerical content—including the cluster catalogue, ages, and dust geometry—is missing. | Merge the two rounds of experimental records and enumerate three issues: the potential variant, reproduction tolerance, and formula exponent. |

| No. | Paper title | DOI / citation | Topic | Main finding | Implicit-knowledge clues | Mining approach |
|---|---|---|---|---|---|---|
| 13 | A dormant overmassive black hole in the early Universe | 10.1038/s41586-024-08210-5<br>Nature, 2024 | JWST · Early black holes | The study identifies a dormant, overmassive black hole in the early Universe. | – The units of the black-hole mass are ambiguous, as are those of the host galaxy's stellar mass.<br>– Whether the paper's actual processing method uses a virial workflow remains to be verified.<br>– One interpretation had access to only a limited portion of the main text because of differences in material access.<br>– Data described as "expected to be public" represent only an eligibility expectation, not verified availability.<br>– The luminosity of ~1e45 is the "corrected" target; the corrected-versus-uncorrected luminosity chain is ambiguous and is further affected by text-extraction problems. | Merge the two rounds of experimental records and enumerate four issue classes: unit ambiguity, reproducibility, simulation availability, and the luminosity chain. |

## Acknowledgements

We sincerely thank Anna for her invaluable help and support throughout the research and the writing of this paper.

## Funding

This work is supported by the Zhejiang Province Key Research and Development Plan (Grant No. 025SSYS0004)

## Figure Credits and Permissions

Figures 1–8 were created by the authors for this study and contain no third-party copyrighted material. Any values or qualitative findings drawn from the reproduced papers are attributed through the text and References; no copyrighted figure panels were reproduced. Figures are vectorized or rendered at a resolution that remains legible after the scaling applied in arXiv/PDF production.

## Use of Generative AI

During the preparation of this work, the authors used large language models, including the 021 Science Foundation Model, and an LLM-based multi-agent system to assist with code drafting, retrieval and organization of literature, execution of the reproduction experiments reported herein, and language editing of the manuscript. All scientific questions, methodological decisions, audit classifications, interpretation of results, and verification of the analysis were performed by the authors, who take full responsibility for the content of this article.